\documentclass[%
reprint,
amsmath,amssymb,
aps,
pra,
]{revtex4-1}

\usepackage{graphicx}% Include figure files
\usepackage{subfigure}
\usepackage{dcolumn}% Align table columns on decimal point
\usepackage{bm}% bold math
\begin{document}

	%\preprint{APS/123-QED}
	
	\title{Enhancing near-field heat transfer between composite structures through strongly-coupled surface modes}% Force line breaks with \\

	\author{W. B. Zhang, C. Y. Zhao}	
	\email{changying.zhao@sjtu.edu.cn}
	\author{B. X. Wang}
	\affiliation{%
		Institute of Engineering Thermophysics, Shanghai Jiao Tong University, Shanghai, 200240, China
	}%

	\begin{abstract}
		In this work, we study the near-field heat transfer between composite nanostructures. It is demonstrated that thermally excited surface plasmon polaritons, surface phonon polaritons, and hyperbolic phonon polaritons in such composite nanostructures significantly enhance the near-field heat transfer. To further analyze the underlying mechanisms, we calculate energy transmission coefficients and obtain the near-field dispersion relations. The dispersion relations of composite nanostructures are substantially different from those of isolated graphene, silicon carbide (SiC) films, and SiC nanowire arrays due to the strong coupling effects among surface polaritonic modes. We identify four pairs of strongly coupled polaritonic modes with considerable Rabi frequencies in graphene/SiC film composite structures that greatly broaden the spectral peak. We confirm that near-field strong coupling effects between surface plasmon polaritons and hyperbolic phonon polaritons in the in-plane Reststrahlen band are different from those in the out-of-plane Reststrahlen band due to the different types of hyperbolicity. In addition, we analyze the effective tunability of the near-field heat transfer of graphene/SiC nanowire arrays composite structures by adjusting the chemical potential of graphene, the height and volume filling factor of the SiC nanowire arrays. This work provides a method to manipulate the near-field heat transfer with the use of strongly coupled surface polaritonic modes.
		
	\end{abstract}
	
	\pacs{Valid PACS appear here}% PACS, the Physics and Astronomy
	\maketitle
	
	%\tableofcontents
	
	\section{\label{sec:level1} Introduction}
	Because of the tunneling effect of evanescent waves, near-field thermal radiation can exceed the blackbody radiation limit, which is administrated by the well-known Planck's Law, by several orders of magnitude \cite{polder1971theory,volokitin2007near,mulet2002enhanced}, especially when surface plasmon polaritons (SPPs) or surface phonon polaritons (SPhPs) are excited \cite{liu2014near,chalabi2015near,dai2015enhanced,shi2013tuning}. The enhancement and further manipulation of near-field heat transfer (NFHT) show promising wide-range applications, such as near-field thermophotovoltaics \cite{svetovoy2014graphene,Park2008Performance,ilic2012overcoming,zhao2017high}, noncontact refrigeration \cite{chen2015heat,guha2012near,chen2016near}, near-field nanoimaging \cite{dai2015subdiffractional,kawata2009plasmonics,fei2012gate}, and information processing \cite{Philippe2014Near,kubytskyi2014radiative}.
	
	To achieve higher enhancement of NFHT, various types of surface polaritonic modes have been extensively studied for their ability to enhance photon tunneling \cite{messina2017radiative,yang2013radiation,wang2017thermoradiative,fu2006nanoscale,miller2014effectiveness}. Recently, graphene is demonstrated to be a good candidate to support SPPs for excellent tunability from near-infrared to terahertz frequencies, owing to the possibility of electrostatic doping \cite{Bokdam2011Electrostatic} and its strong ability to produce higher confinement and lower losses compared to metals \cite{jablan2009plasmonics}. It has been shown that thermally excited graphene SPPs can strongly mediate the NFHT between graphene sheets \cite{ilic2012near,abbas2017strong}. When structures consist of graphene with other dielectric materials or metamaterials, the NFHT can be further enhanced due to the coupling effects among various surface polaritonic modes \cite{messina2017graphene,zhao2017active,liu2014graphene,basu2015near}. And if the interaction between surface polaritonic modes is strong enough, a hybrid mode will form due to the strong coupling effects. For example, hyperbolic phonon polaritons (HPPs) supported by hyperbolic materials can couple with the SPPs of graphene to form new hybrid modes, resulting in nearly perfect photon tunneling \cite{yin2016super,hajian2017hybrid,shi2015near,biehs2017near,messina2016hyperbolic}. Researchers have recently studied graphene/hexagonal boron nitride (hBN) multilayer heterostructures for NFHT and demonstrated an infinite limit \cite{shi2017enhanced,zhao2017enhanced}. Because the coupled SPP-HPP hybrid modes in graphene/hBN heterostructures suffer little from Ohmic losses, their propagation length is 1.5 to 2.0 times greater than that of HPPs in hBN \cite{dai2015graphene}. What's more, the near-field effects can also be used to largely enhance heat transfer between far separated objects \cite{asheichyk2017heat,asheichyk2018heat,dong2018long,messina2018surface}. 
	In addition, the strong coupling effects have been investigated for local density of states in graphene-covered systems \cite{messina2013tuning,zhao2015strong,liu2015highly}, which can significantly reduce the losses in light propagation. However, the role of these strong coupling effects in the NFHT remains unclear. Unlike the coupling in far-field region, the strong coupling effects between near-field nanostructures are sensitive to the near-field gap distance. To further understand and control the NFHT in these near-field systems, the mechanisms of excitation, coupling, and interference of various frequency-resonant modes (FRMs), which stand for branches of surface polaritonic modes, must be comprehensively explored.
	
	In this work, we investigate the strong coupling effects on NFHT by studying the NFHT of two composite structures: the graphene/silicon carbide (SiC) film composite structure and the graphene/SiC nanowire arrays composite structure. The surface polaritonic modes of these structures are excited in the infrared region. To further analyze the underlying mechanisms, we calculate the energy transmission coefficients and obtain the near-field dispersion relations. Due to the strong coupling effects between different polaritonic modes, the dispersion relations of composite structures differ substantially from those of isolated graphene, SiC films, and SiC nanowire arrays. Therefore, we further identify strongly coupled polaritonic modes with considerable Rabi frequencies in composite structures. In addition, we analyze the contributions of the composite structures' parameters, such as the chemical potential of graphene, the height and volume filling factor of SiC nanowire arrays, to the near-field strong coupling effects. As a result, broadband and tunable NFHT can be achieved by manipulation of the strong coupling effects. This work provides a method to control NFHT by strongly coupled surface polaritonic modes, which shows promise in near-field applications.
	
	\section{\label{sec:level2} Theory}
	\subsection{\label{sec:level21}Model of NFHT between composite structures}
	
	Fig. 1 illustrates the schematics of the NFHT between graphene/SiC film composite structures and the NFHT between graphene/SiC nanowire arrays composite structures. Based on fluctuational electrodynamics using dyadic Green’s functions, the heat flux between two structures can be calculated by \cite{liu2014application}:
	\begin{equation}
	\begin{split}
    \begin{array}{l}Q=\frac1{4\pi^2}\int_0^\infty\left[\Theta\left(T_E,\omega\right)-\Theta\left(T_R,\omega\right)\right]\\\;\;\;\;\;\;\times\;\left[\int_0^\infty k_\perp\underset{j=s,p}{\sum\;}\xi_j\left(\omega,k_\perp\right)\operatorname d k_\perp\right]\operatorname d\omega\end{array},
	\end{split}
	\end{equation}
	where the local thermal equilibrium temperatures ${T_E}$ and ${T_R}$ are identified as the emitter and receiver, respectively. $k_\perp$ is the transverse wavevector of the thermal radiation waves, and $j=s,p$ stands for $s$- or $p$-polariton modes, respectively. $\Theta\left(T,\omega\right)=\hbar\omega/\left[\exp\left(\hbar\omega/k_BT\right)-1\right]$ is the mean energy of the thermal harmonic oscillators.
	
	\begin{figure}[htbp]
		\centering
		\includegraphics[width=8.6cm]{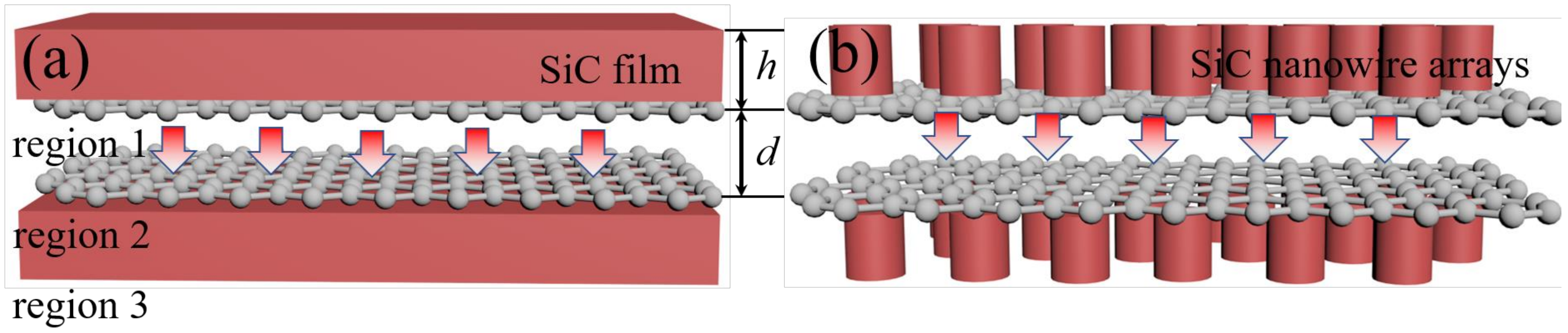}
		\caption{Schematic of (a) NFHT between graphene/SiC film composite structures and (b) NFHT between graphene/SiC nanowire arrays composite structures.}
	\end{figure}
	The energy transmission coefficient $\xi_j\left(\omega,k_\perp\right)$ is given by \cite{ito2017dynamic,ikeda2017tunable,muller2017many}:
	
	\begin{equation}
    \xi_j\left(\omega,k_\perp\right)=\left\{\begin{array}{lc}\frac{\left(1-\left|r_{j,E}\right|^2\right)\left(1-\left|r_{j,R}\right|^2\right)}{\left|1-r_{j,E}r_{j,R}e^{2ik_{z0}d}\right|^2},&k_\perp<k\\\frac{4\mathrm I\mathrm m\left(r_{j,E}\right)\mathrm I\mathrm m\left(r_{j,R}\right)e^{-2\left|k_{z0}\right|d}}{\left|1-r_{j,E}r_{j,R}e^{2ik_{z0}d}\right|^2},&k_\perp>k\end{array}\right.,
	\end{equation}
	where $r_{j,E}$ and $r_{j,R}$ indicate the Fresnel reflection coefficients of the emitter and receiver, respectively. $d$ indicates the vacuum gap distance. $k=\omega/c$ means the magnitude of the wavevector and $k_{z0}=\left(k^2-k_\perp^2\right)^{1/2}$ denotes the $z$-component of the wavevector in vacuum. Note that in Eq. (2), the expressions of energy transmission coefficients for propagating waves ($k_\perp<k$) and evanescent waves ($k_\perp>k$) are different.
	
	The Fresnel reflection coefficients for the graphene-covered composite structures are shown in the following forms \cite{saarinen2008green}:
	\begin{equation}
	\label{E3}
	\begin{array}{rcl}r_{p,l}&=&\frac{r_{12,p}+\left(1-r_{12,p}-r_{21,p}\right)r_{21,p}e^{2ik_{z,2}h}}{1-r_{21,p}r_{23,p}e^{2ik_{z,2}h}}\end{array},
	\end{equation}
	\begin{equation}
	\begin{array}{rcl}r_{s,l}&=&\frac{r_{12,s}+\left(1+r_{12,s}+r_{21,s}\right)r_{21,s}e^{2ik_{z,2}h}}{1-r_{21,s}r_{23,s}e^{2ik_{z,2}h}}\end{array},
	\end{equation}
	where indexes 1, 2, and 3 refer to the vacuum region above the composite structure, in the composite structure, and the vacuum region below the composite structure, respectively, as shown in Fig. 1. In addition, the $l$ = $E$ or $R$ stands for the emitter or receiver, and $h$ indicates the thickness of the SiC film or the height of the SiC nanowires. Eq. (3) and Eq. (4) can also be used for nanostructures with only monolayer graphene by setting $r_{23}=0$.
	
	For a dielectric material covered by monolayer graphene, the Fresnel reflection coefficients become \cite{messina2013graphene}:
	\begin{equation}
	\label{E4}
	\begin{array}{rcl}r_{ab,p,l}&=&\frac{k_{z,a}\varepsilon_{\perp,b}-k_{z,b}\varepsilon_{\perp,a}+\sigma k_{z,a}k_{z,b}/\left(\varepsilon_0\omega\right)}{k_{z,a}\varepsilon_{\perp,b}+k_{z,b}\varepsilon_{\perp,a}+\sigma k_{z,a}k_{z,b}/\left(\varepsilon_0\omega\right)}\end{array},
	\end{equation}
	\begin{equation}
	\begin{array}{rcl}r_{ab,s,l}&=&\frac{k_{z,a}-k_{z,b}-\sigma\mu_0^2\omega}{k_{z,a}+k_{z,b}+\sigma\mu_0^2\omega}\end{array},
	\end{equation}
	where $k_{z,n}=\left(\varepsilon_{\perp,n}k^2-\varepsilon_{\perp,n}k_\perp^2/\varepsilon_{\parallel,n}\right)^{1/2}$ with $n$ = 1, 2, or 3 in Eq. (3)$-$(6). $\varepsilon_\perp$ and $\varepsilon_\parallel$ are the vertical and parallel components of the relative dielectric tensor, respectively. If a monolayer of graphene is present between media $a$ and $b$, then $a$ = 1, 2 and $b$ = 1, 2 or 3. The role of the monolayer graphene can be regarded as a current sheet conductivity ($\sigma$). If only SiC film or SiC nanowire arrays are present, $\sigma$ in Eq. (5) and Eq. (6) should be set to be zero. Note that $\sigma$ is the extensively admitted conductivity of graphene at temperature $T_l$ modeled by the local random phase approximation (LRPA) \cite{falkovsky2007space},
	\begin{equation}
	\label{E5}
	\begin{array}{l}\sigma=\frac{2ie^2k_BT_lIn\left[2\mathrm c\mathrm o\mathrm s\mathrm h\left[\mu/\left(2k_BT_l\right)\right]\right]}{\left(\omega+i/\tau\right)\pi\hslash}\\\;\;\;\;\;\;\;\;+\frac{e^2}{4\hslash^2}\left[f\left(\frac{\hslash\omega}2\right)+\frac{i4\hslash\omega}\pi I\right],\end{array}
	\end{equation}
	where the relaxation time of graphene at 300 K is about $\tau=100\;fs$ \cite{Abajo2014Graphene}. In Eq. (7), $f\left(\delta\right)=1/\left[\left[{\rm{cosh}}\left(\mu/k_BT_l\right)/{\rm{sinh}}\left(\delta/k_BT_l\right)\right]+{\rm{coth}}\left(\delta/k_BT_l\right)\right]$, and $I=\int_0^\infty\left[f\left(\delta\right)-f\left(\hbar\omega/2\right)\right]/\left[\left(\hbar\omega\right)^2-4\delta^2\right]\operatorname d\delta$. The conductivity of graphene consists of the intraband (the first term) and interband (the second term) contributions of electron-photon scattering processes. The above expression is widely adopted \cite{Minovkoppensich2011Graphene} and has been verified by considerable experimental data \cite{Chen2013Strong,Fang2013Gated}.
	
	SiC film is a nonmagnetic polar material. According to the Lorentz model, the dielectric function of SiC is given by \cite{smith1985handbook}:
	\begin{equation}
	\label{E6}
	\varepsilon_{SiC}=\varepsilon_\infty\frac{\omega^2-\omega_{LO}^2+i\Gamma\omega}{\omega^2-\omega_{TO}^2+i\Gamma\omega}.
	\end{equation}
	where $\omega_{LO}$ and $\omega_{TO}$ are the longitudinal optical phonon frequency and the transverse optical phonon frequency, respectively.
	
	SiC nanowire arrays can be regarded as a homogenous medium that possesses anisotropic optical properties, thus the uniaxial tensor of SiC nanowire arrays can be expressed as:
	\begin{equation}
	\widehat\varepsilon=\begin{pmatrix}\varepsilon_\parallel&0&0\\0&\varepsilon_\parallel&0\\0&0&\varepsilon_\perp\end{pmatrix}
	\end{equation}
	
	According to the effective medium theory (EMT), which is frequently used to predict the NFHT between gratings \cite{Biehs2011Modulation} and between multilayers \cite{Zhang2017Validity}, the dielectric function of the parallel and vertical components of SiC nanowire arrays are provided by \cite{Biehs2012Hyperbolic}:
	\begin{equation}
	\varepsilon_\parallel=\frac{\varepsilon_{SiC}\left(1+f\right)+1-f}{\varepsilon_{SiC}\left(1-f\right)+1+f},
	\end{equation}
	\begin{equation}
	\varepsilon_\perp=\varepsilon_{SiC}f+1-f,
	\end{equation}
	where $f$ is the volume filling factor. A change in this volume filling factor can affect the hyperbolic frequency of the dielectric function. To control for the HPPs of SiC nanowire arrays excited in the infrared region, we set $f$ = 0.2.

	Note that calculation of the volume filling factor with the EMT is only satisfied under the condition in which the characteristic dimension of the nanowires ($d_0$) is much smaller than the wavelength ($\lambda$) of the radiation. The model prediction and the simulation for doped Si nanowire arrays have been compared \cite{liu2013wideband,liu2014application}. It was confirmed that the EMT model can accurately predict the properties of doped Si nanowire arrays with a small volume filling factor ($f<0.6$) which is within the wavelength region of interest ($d_0\ll\lambda$). What's more, using the rigorous coupled-wave analysis (RCWA) approach \cite{yang2016spectrally}, which is described in appendix, we find that both methods yield essentially identical results with less than 0.5\% in the predicted total heat flux of the graphene/SiC nanowire arrays composite structures when the gap distance is more than 50 nm. Therefore, the volume filling factor of SiC nanowire arrays used in this work is reliable for the EMT model.
	
	\section{\label{sec:level3} Results and Discussion}
	\subsection{\label{sec:level31}Mechanisms for enhancement of NFHT between composite structures}
	In this work, the emitter and receiver thermal equilibrium temperatures are $T_E = 310 \;\rm{K}$ and $T_R = 290 \;\rm{K}$, respectively. As illustrated in Fig. 2a, the total heat flux between graphene/SiC film composite structures is the highest ($344.5 \;\rm{kW/m^2}$) when the gap distance is 50 nm. Due to the exponential decay feature of large-wave-vector evanescent waves, the heat flux of graphene/SiC film composite structures decreases more rapidly than that of monolayer graphene structures. When the gap distance is about 5000 nm, the heat fluxes of five structures are nearly the same, approaching the far-field region. The heat flux of the graphene/SiC nanowire arrays is lower than that of the graphene/SiC film at the narrow vacuum gap distance because the new hybrid modes, such as SPP-SPhP hybrid modes and SPP-HPP hybrid modes, formed on the composite structures. However, different hybrid modes have different effects on the NFHT, and the details are discussed below.
	
	Mechanisms that enhance the NFHT between composite structures can be explained by the contours of energy transmission coefficients. In this work, the surface polaritonic modes can only be excited by TM waves because $\xi_p\gg\xi_s$. Fig. 2b-f shows the contours of the energy transmission coefficients for various structures. The transverse wavevector is normalized by using $k_0=\omega_0/c$ with $\omega_0=1\times10^{14}\;\rm{rad/s}$. The bright bands presented in Fig. 2 demonstrate the higher photon tunneling rate that results from the excitation of different surface polaritonic modes. According to Fig. 2b, the two bright bands separate in the low-frequency region and converge in the high-frequency region, which represent coupled SPPs between the two monolayer graphene structures. Fig. 2c presents bright bands that represent thermally excited SPhPs near the frequency of the SiC-vacuum interface \cite{Gubbin2017Theoretical}. The enlarged dispersion relation in the inset of Fig. 2c shows that four FRMs converge in the high-transverse-wave-vector region. The horizontal white dotted line indicates the frequency of the SiC-vacuum interface, and the horizontal green dotted-dashed lines indicate the longitudinal optical phonon frequency and the transverse optical phonon frequency, respectively. These three lines show the limits of the near-field dispersion relation of the SiC film structures. In the thin film structures, the evanescent field of the SPhPs linked to each interface can interact with each other, which leads to division of the SPhPs dispersion relation into antisymmetric FRMs and symmetric FRMs. The symmetric FRM corresponds to the case in which the tangential electric field has a symmetric distribution, and vice versa for the antisymmetric FRM \cite{dionne2005planar}. Fig. 2d shows that the HPPs of the SiC nanowire arrays are thermally excited in each Reststrahlen band ($1.494\times10^{14}\;\rm{rad/s}\;\sim\;1.708\times10^{14}\;\rm{rad/s}$ and $1.771\times10^{14}\;\rm{rad/s}\;\sim\;1.797\times10^{14}\;\rm{rad/s}$). The heat flux of the SiC nanowire arrays (Fig. 2d) is higher than that of the SiC film (Fig. 2c), which is consistent with Fig. 2a, because the electromagnetic wavevector in the hyperbolic-frequency region can theoretically approach infinity and the photon density becomes divergent, which can strengthen the electromagnetic interaction.
	\begin{figure}[htbp]
		\centering
		\includegraphics[width=8.6cm]{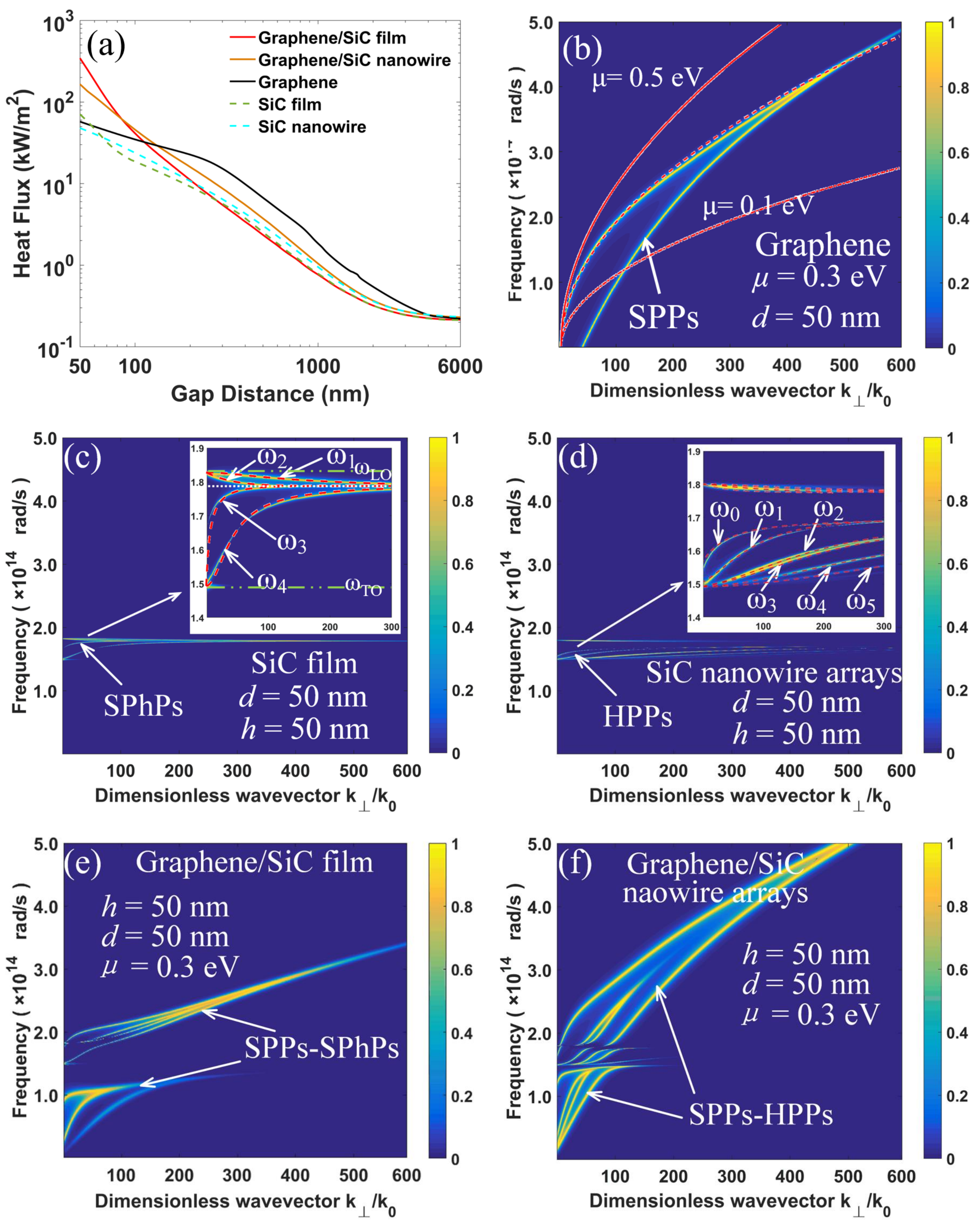}
		\caption{(a) Heat flux versus vacuum gap distance $d$ from 50 to 6000 nm for various structures. Contours of energy transmission coefficients for various structures: (b) monolayer graphene, (c) SiC film, (d) SiC nanowire arrays, (e) graphene/SiC film composite structure, and (f) graphene/SiC nanowire arrays composite structure. The chemical potential of graphene is $\mu$ = 0.3 eV. Thickness of SiC film or height of SiC nanowires is $h$ = 50 nm.}
	\end{figure}
	
	However, when the SiC film is covered by a monolayer graphene, the heat flux between the graphene/SiC film composite structures is higher than that of the SiC nanowire arrays at the narrow gap distance. The reason is shown in Fig. 2e, in which bright bands split into two branches because the SPhPs excited in the isotropic polar materials (i.e., the SiC film) couple with the SPPs in graphene to form SPP-SPhP hybrid modes. Similar phenomena were observed in the coupling between SPPs in a graphene or thin metal layer with the SPhPs in $\rm{SiO_2}$ substrates \cite{messina2013tuning}. However, the SPhP modes excited near $1.786\times10^{14}\;\rm{rad/s}$ (i.e., the frequency of the SiC vacuum interface) are attenuated appreciably relative to those in Fig. 2c. The SPhPs nearly disappear in the high-transverse-wave-vector region. This phenomenon is also a manifestation of the strong coupling effects between the graphene/SiC film composite structures, which can induce a photonic gap without any modes. Fig. 2f obviously shows hybrid mode flattening when the SPPs' approach is near two hyperbolic Reststrahlen bands; this is a result of the splitting effect or mode repulsion between the SPPs in graphene and the HPPs in the SiC nanowire arrays. This effect allows the two hyperbolic Reststrahlen bands to break the otherwise continuous SPPs into three frequency regions (below $1.494\times10^{14}\;\rm{rad/s}$, between $1.494\times10^{14}\;\rm{rad/s}$ and $1.797\times10^{14}\;\rm{rad/s}$, and above $1.797\times10^{14}\;\rm{rad/s}$), which enables a high state density to occur at some band edges and boosts photon tunneling. The SPPs-HPPs hybrid mode of the excited region of the graphene/SiC nanowire arrays composite structures is larger than that of the graphene/SiC film composite structures in the frequency direction and in the transverse-wavevector direction. However, it has a larger photonic gap in Fig. 2e, and the energy transmission coefficients of the graphene/SiC film composite structures are higher than those of the graphene/SiC nanowire arrays composite structures due to the strong coupling effects. Thus, the heat flux of the graphene/SiC film composite structures still exceeds that of the graphene/SiC nanowire arrays composite structures when the vacuum gap distance is narrow.
	
	\subsection{\label{sec:level32}Asymptotic analysis of surface polaritonic modes}
	To further analyze the origins and consequences of the strong coupling effects, we first perform asymptotic analysis of the near-field dispersion relation. According to Eq. (2), the radiative heat flux diverges when the following condition is fulfilled:
	\begin{equation}
	1-r_{j,E}r_{j,R}e^{2ik_{z0}d}=0.
	\end{equation}
	
	We can obtain each FRM by solving Eq. (12). 
	
	In terms of the dielectric material such as SiC film with isotropic optical properties, for a surface wave with $k_\perp\gg k$, the $z$-component of the wavevector is $k_{z,n}=\left(\varepsilon_{\perp,n}k_0^2-\varepsilon_{\perp,n}k_\perp^2/\varepsilon_{\parallel,n}\right)^{1/2}$ with $n$ = 1, 2, or 3, which can be approximated by $k_{z,n}\approx i k_\perp$. Ignoring the imaginary part of Eq. (12), we obtain $r_{12}=S\left(k_\perp,h,d\right)$, where $S$ is satisfied with the following conditions:
	\begin{equation}
	S\left(k_\perp,h,d\right)=\pm\sqrt{\frac{-b\pm\sqrt{b^2-4ac}}{2a}},
	\end{equation}
	\begin{equation}
	\left\{
	\begin{aligned}
	\begin{array}{l}a=e^{k_\perp\left(d-2h\right)}\\b=2e^{-k_\perp d}-2e^{k_\perp d}-e^{-k_\perp\left(d+2h\right)}-e^{-k_\perp\left(d-2h\right)}\\c=e^{k_\perp\left(d+2h\right)}\end{array}.
	\end{aligned}
	\right.
	\end{equation}
	
	By using the dielectric function of SiC film in Eq. (8), the FRMs of SiC film can be obtained as:
	\begin{equation}
	\begin{array}{l}\omega_S^\pm\approx\left[\frac{S\left(\varepsilon_\infty\omega_{LO}^2+\omega_{TO}^2\right)+\left(\omega_{TO}^2-\varepsilon_\infty\omega_{LO}^2\right)}{S\left(\varepsilon_\infty+1\right)+\left(1-\varepsilon_\infty\right)}\right]^{1/2}\end{array},
	\end{equation}
	where the losses are small and thus ignored. Eq. (15) offers an approximation of the four FRMs of the two film structures as a function of $d$, $h$, and $k_\perp$. The resonant frequencies at which the radiative heat flux between two SiC film structures reaches its maximal value can be determined from Eq. (15), shown as the red dashed lines in the inset of Fig. 2c. The analytical formulas can accurately predict each FRM. $\omega_1$ corresponds to the highest FRM, and $\omega_4$ is consistent with the lowest FRM. According to the FRMs in the dispersion relation, numerous electromagnetic states are available within a narrow spectral band near the frequency of the SiC-vacuum interface, as illustrated in Fig. 4a. We can therefore assume that the radiative heat flux is maximal at $k_\perp$ in which $\left|d k_\perp/d\omega\right|$ is the highest. Fig. 3a presents the SPhPs near-field dispersion relation of SiC films of various thicknesses. Comparison of the dispersion relations for 20-, 50- and 100-nm-thick SiC films in a vacuum shows that the splitting of the resonance becomes more pronounced with a thinner film. For the large dimensionless wavevector, the dispersion relations of both symmetric and antisymmetric modes approach asymptotically the dispersion curve of a single SiC-vacuum interface. Indeed, the penetration depth of the SPhPs in the film is small for the large dimensionless wavevector. Therefore, when the dimensionless wavevector is large, the antisymmetric and symmetric branches degenerate because the SPhPs do not couple inside the films, so the FRMs of SPhPs at each interface behave independently of each other \cite{economou1969surface}.
	\begin{figure}[htbp]
		\centering
		\includegraphics[width=8.6cm]{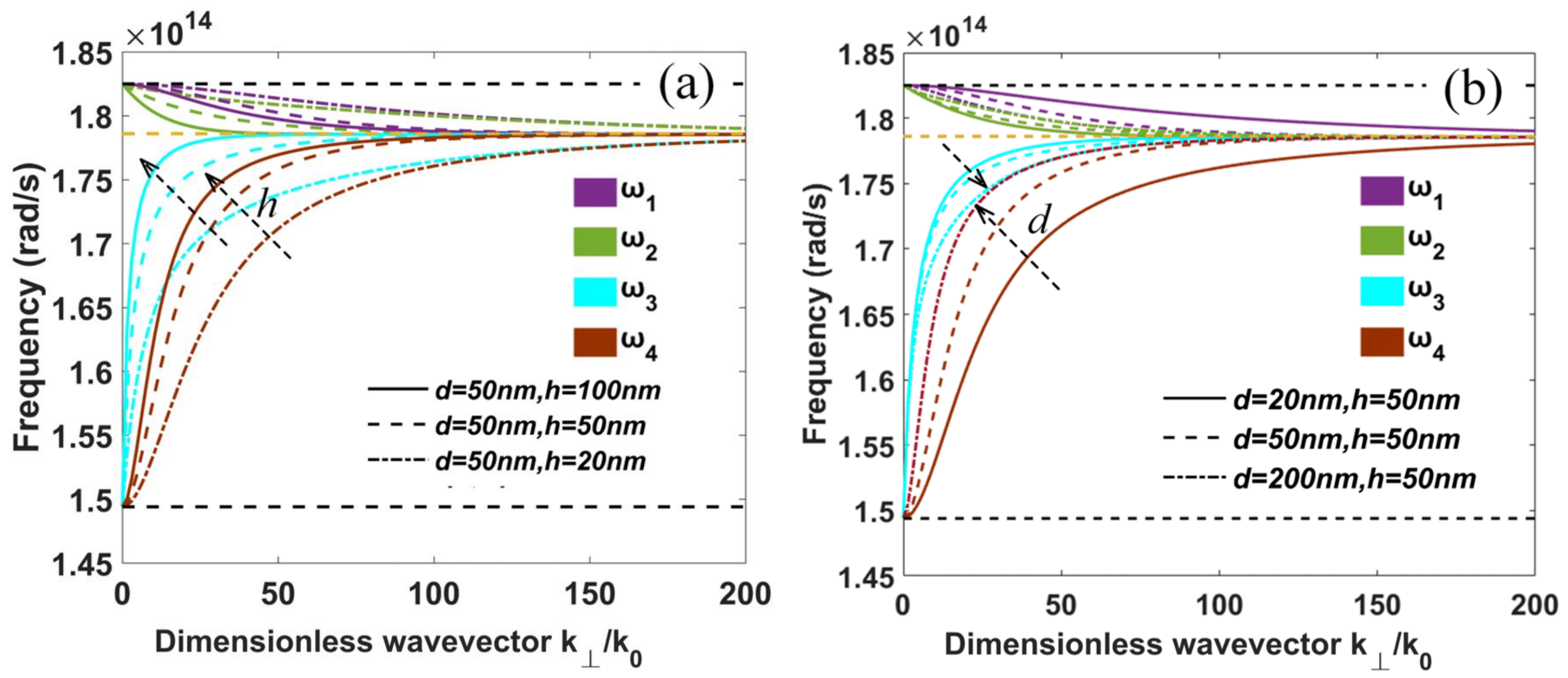}
		\caption{SPhPs near-field dispersion relations for SiC film structures with (a) different thicknesses and (b) different vacuum gap distances.}
	\end{figure}

	When two SiC films are placed in close proximity, further coupling occurs, and the dispersion relation splits into four FRMs that show the antisymmetric and symmetric resonances for the entire structure. This is illustrated in Fig. 3b by analysis of the cases $d$ = 20, 50 and 200 nm, in which all films are 50 nm thick. As the vacuum gap distance increases, $\omega_1$ and $\omega_2$ ($\omega_3$ and $\omega_4$) grow closer to each other, which weakens the coupling between the SPhPs of the two SiC films. If the gap distance is sufficiently large, the SPhPs on each film behave independently, and only two FRMs exist in the dispersion relation.
	
	However, in terms of hyperbolic metamaterials such as SiC nanowire arrays with anisotropic optical properties, for a surface wave with $k_\perp\gg k$, the $z$-component of the wavevector can be estimated by $k_{z,n}\approx i\sqrt{\varepsilon_\perp/\varepsilon_\parallel} k_\perp=\varphi k_\perp$. Eq. (12) enables us to further simplify the expression of $r_p$. If dissipation is neglected, $\varepsilon_\perp$ and $\varepsilon_\parallel$ are both real, and the allowable value of momentum $k_{z,n}$ obeys the Fabry-Perot condition \cite{kumar2015tunable,dai2014tunable}:
	\begin{equation}
	\delta_1+\delta_2+2k_{z.n}h=2m\pi,
	\end{equation}
	where $\delta_1$ and $\delta_1$ are the phases of the reflection coefficient. Moreover, $\tan\delta_1=\tan\delta_1=\varphi/\varepsilon_\parallel$. Therefore, the FRMs of the single SiC nanowire arrays inside the two Reststrahlen bands can be written as:
	\begin{equation}
	k_\perp\left(\omega\right)=-\frac1{\varphi h}\left[m\pi+2arc\tan\left(\frac\varphi{\varepsilon_\parallel}\right)\right],
	\end{equation}
	where integer $m$ = 0, 1, 2, 3, $\cdots$ denotes the resonance order. Eq. (17) has infinite solutions transverse wavevector corresponding to various FRMs of the HPPs. As illustrated in the inset (red dashed lines) of Fig. 2d, $\omega_0$ indicates the lowest-order (i.e., zeroth-order) FRM in the in-plane Reststrahlen band. However, the out-of-plane reststrahlen band has no zeroth-order FRM. In other words, the minimum $m$ equals 1 rather than 0 in the out-of-plane Reststrahlen band because the nature of these FRMs differs completely between the in-plane and out-of-plane Reststrahlen bands due to different kinds of hyperbolicity \cite{huang2013zeroth}. As a result, the in-plane Reststrahlen band possesses several FRMs for $m$ = 0, 1, 2, 3, 4, 5, as illustrated by the red dashed lines in the inset of Fig. 2d. The lower-order FRMs occur in a higher-frequency region in the in-plane Reststrahlen band, but the lower-order FRMs occur in a lower-frequency region in the out-of-plane reststrahlen band because of the hyperbolicity, which indicates a negative in-plane dielectric function, a positive out-of-plane component. In addition, we can predict that the field is generally confined inside the SiC nanowire arrays instead of in the surrounding media or at the interfaces, especially for the higher-order FRMs. According to Eq. (17), the FRMs in the SiC nanowire arrays depend strongly upon the height of the SiC nanowires. As the height of the SiC nanowire arrays increases, more higher-order FRMs of HPPs will be excited in the SiC nanowire arrays, and the FRMs with different orders of HPPs will move closer to the transverse optical phonon frequency, which can strengthen the coupling of HPPs. Fig. 2a (red dashed lines) depicts the dispersion relation of the SPPs for a single monolayer graphene for three selected chemical potentials ($\mu$ = 0.1 eV, 0.3 eV, and 0.5 eV). These parabolic curves can be described by \cite{Wunsch2007Dynamical}:
	\begin{equation}
	k_\perp\left(\omega\right)=\frac{2\hbar\omega^2\eta}{g\alpha\mu V_F},
	\end{equation}
	where $\alpha\equiv e^2/\left(4\pi\kappa_0\hbar V_F\right)\simeq2.5$ \cite{stern1967polarizability}. $g=g_Sg_V=4$, $g_S$ and $g_V$ denote the spin and valley degeneracy, respectively. $\eta$ indicates the high frequency screening. The Fermi velocity $V_F=3at/\left(2\hbar\right)$ is determined by the carbon-carbon distance $a=1.42$ $\AA$ and the nearest neighbour hopping energy $t=2.7$ eV \cite{peres2006electronic}.
	Obviously, with the increase of the chemical potential of graphene, the SPPs cannot occur in the high-frequency region, which is disadvantageous to NFHT enhancement.
	
	\subsection{\label{sec:level33}Strong coupling effects of surface polaritonic modes in composite structures}
	Based on Sections A and B, this subsection further investigates the effects of the coupling modes on NFHT enhancement. As illustrated in Fig. 4a, monolayer graphene, SiC film, and SiC nanowire arrays each show a single narrow peak that corresponds to the thermally excited SPPs, SPhPs, and HPPs, respectively. However, as for the structure of the graphene/SiC nanowire arrays (orange solid line) in Fig. 4a, four peaks exist in the curve of spectral heat flux; two can be attributed to the HPPs (in-plane and out-of-plane Reststrahlen bands) between $1.494\times10^{14}\;\rm{rad/s}$ and $1.797\times10^{14}\;\rm{rad/s}$, and the other two can be attributed to the SPP-HPP hybrid modes, which exist near $1.494\times10^{14}\;\rm{rad/s}$ and $1.797\times10^{14}\;\rm{rad/s}$ due to the strong coupling effects of SPPs and HPPs. The spectral heat flux of the graphene/SiC film composite structure (red solid line) has two peaks, which are generated from the strong coupling of SPPs and SPhPs. The low-frequency peak is at $1.07\times10^{14}\;\rm{rad/s}$, and the high-frequency peak is at $2.23\times10^{14}\;\rm{rad/s}$. These two peak frequencies differ substantially from the optical phonon frequencies of the SiC film, which suggests that strong coupling indeed takes place in graphene/SiC film composite structures. Moreover, the strong coupling effects can greatly broaden the spectral peak.
	
	\begin{figure}[htbp]
		\centering
		\includegraphics[width=8.6cm]{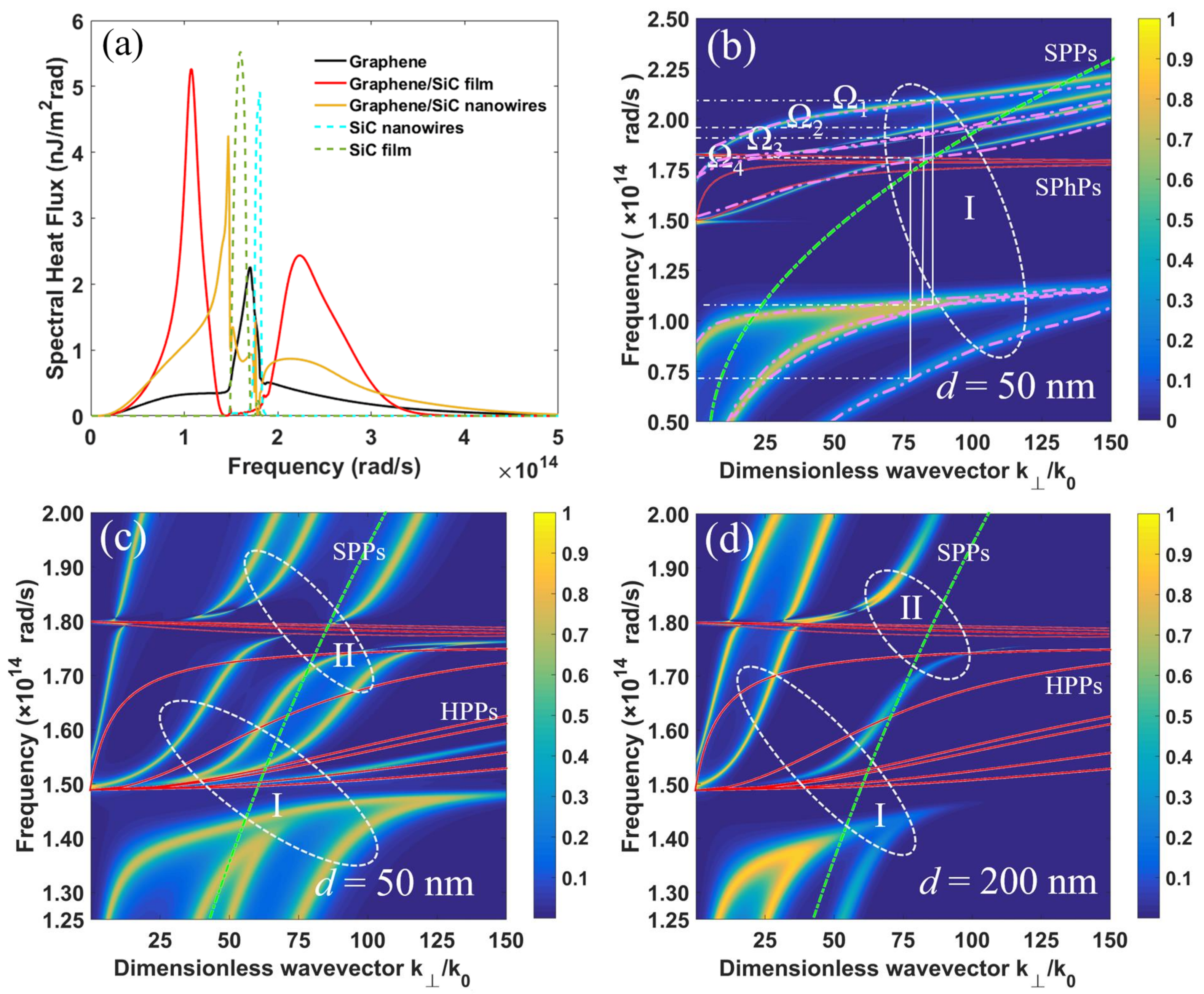}
		\caption{ (a) Spectral heat flux of various structures with a gap distance of 50 nm. Contours of energy transmission coefficients for (b) graphene/SiC film composite structures with a gap distance of 50 nm, (c) graphene/SiC nanowire arrays composite structures with a gap distance of 50 nm, and (d) graphene/SiC nanowire arrays composite structures with a gap distance of 200 nm.}
	\end{figure}

	The strong coupling regime can be defined as the splitting being large enough compared to the linewidths of the coupled states. Analytically, the strong coupling can be described by the system of the two coupled oscillators, modeling the SPP and SPhP modes, respectively. The eigenfrequencies $\omega^\pm$ of the coupled system are then given by \cite{rudin1999oscillator}: 
	\begin{equation}
	\omega^\pm=\frac{\omega_{\rm{S}}+\omega_{\rm{G}}\pm\sqrt{\left(\omega_{\rm{S}}-\omega_{\rm{G}}\right)^2+\Omega^2}}2,
	\end{equation}
	where $\omega_{\rm{S}}$ and $\omega_{\rm{G}}$ are the frequencies of the uncoupled SPP and SPhP modes and they can be obtained by Eq. (15) and Eq. (18). $\Omega$ is the Rabi frequency as a fitting parameter, quantifying
	the mutual interaction. Fig. 4b explicitly shows the strong coupling effects between the SPP and SPhP modes, in which the two splitted dispersive branches are shown as two bright bands. The four red lines denote the FRMs of SPhPs given by Eq. (15), and the parabolic green dotted-dashed line denotes the graphene SPPs given by Eq. (18). By employing Eq. (19) to fit the dispersion analytically, we can determine four strong coupling Rabi frequencies: $\Omega_1=1.01\times10^{14}\;\rm{rad/s}$, $\Omega_2=8.6\times10^{13}\;\rm{rad/s}$, $\Omega_3=8.3\times10^{13}\;\rm{rad/s}$, and $\Omega_4=1.11\times10^{14}\;\rm{rad/s}$, which denote the four pairs of strongly coupled modes. And the pink dotted-dashed lines in Fig. 4b are the fitting results. Therefore, it is those strongly coupled polaritonic modes that significantly enhance the NFHT. According to the definition of the strong coupling regime, the strong coupling condition requires that the energy exchange rate between the two strongly coupled oscillators should exceed the loss rate, resulting in the appearance of two distinct frequencies in the spectrum \cite{auffeves2010controlling,torma2014strong} (as shown in Fig. 4a). A reliable and sufficient condition of the strong coupling is thus the splitting occurring if $\Omega/\gamma>1$ is fulfilled, where $\gamma=\left(\gamma^{\rm{SPPs}}+\gamma^{\rm{SPhPs}}\right)/2$ is the average of the loss rates of the two modes. As for graphene/SiC film composite structure, the losses of the SPP and SPhP modes are small. We can demonstrate that the strong coupling effects observed in the near-field system are correct (While $\gamma^{\rm{SPhPs}}$ can be readily calculated using the transfer matrix method, detemination of $\gamma^{\rm{SPPs}}$ is not straightforward. As an estimation, we assume the proportionality of the loss rate to the imaginary part of the permittivity $\rm{Im}\left(\varepsilon\right)$ of the corresponding material where the mode is largely localized.). What's more, we can confirm that the FRMs of SPhPs couple first with each other and then coupled SPhPs couple with the SPPs of the monolayer graphene because four Rabi frequencies exist.
	
	When SiC nanowire arrays is covered by a monolayer of graphene, an interesting phenomenon is observed in which the four lower-order FRMs of HPPs are strongly coupled with the graphene SPPs in the two Reststrahlen bands, as illustrated in regions I and II of Fig. 4c. However, the higher-order FRMs are rarely affected by these coupling effects because of the symmetry mismatch. Fig. 4d illustrates that strong coupling effects still exist with a vacuum gap distance of 200 nm. Compared with Fig. 4c, the main difference is that the coupling effect of the fourth-order FRM disappeared due to its smaller penetration depth. Therefore, we conclude that the strong coupling effects are mainly caused by the SPPs and HPPs of the composite structures.
	
	\subsection{\label{sec:level34}Effects of parameters of composite structures on NFHT}
	We further investigate the NFHT of graphene/SiC nanowire arrays composite structures with various chemical potentials. When the vacuum gap distance is less than 200 nm, the heat flux between the two structures decreases as the chemical potential increases, as illustrated in Fig. 5a. Fig. 5b-d depicts the contours of the energy transmission coefficients of chemical potential $\mu$ = 0.1 eV, 0.3 eV, and 0.5 eV, respectively, with a gap distance of 50 nm. The strong coupling effects at $\mu$ = 0.1 eV are stronger than those at $\mu$ = 0.3 and 0.5 eV. And the strong coupling effects of the higher-order FRMs of HPPs clearly become weaker as the chemical potential increases. The reason is that the chemical potential of graphene directly relates to the electric gate \cite{Grigorenko2012Graphene}, the lower the chemical potential of graphene, the higher the wavevector space of the corresponding resonance mode. In addition, the highest heat transfer level can be obtained at around 0.1 eV rather than at 0 eV because interband transitions play a dominant role in the infrared region at $\mu$ = 0 eV and graphene no longer supports SPPs in this region \cite{Zhao2015Resonance}. However, as the vacuum gap distance increases, the heat flux of the composite structures with lower chemical potential decreases more rapidly than the heat flux of the structures with higher chemical potential due to the exponential decay feature of large-wavevector evanescent waves.
	
	Eq. (3) and Eq. (4) show that the height and volume filling factor of the SiC nanowire arrays are also important to the NFHT between graphene/SiC nanowire arrays composite structures. Fig. 6a shows the total heat flux of the graphene/SiC nanowire arrays, and the volume filling factors are set from 0.1 to 0.4 to confirm the reliability of the EMT model. The heights of the SiC nanowire arrays are set from 20 to 200 nm. Note that if the SiC nanowire arrays exceeded 200 nm, the NFHT enhancement would barely be affected. We can see the highest heat flux near the position of $h$ = 60 nm and $f$ = 0.25 at a vacuum gap distance $d$ = 50 nm, as illustrated in Fig. 6a near point H, because the HPPs of the SiC nanowire arrays can be well excited in the region near this point, which is the same as the SPPs' excited region. In addition, a decrease in the height of the SiC nanowire arrays means that the HPPs can couple well with the SPPs of graphene to form the hybrid HPP-SPP modes.

   \begin{figure}[htbp]
	\centering
	\includegraphics[width=8.6cm]{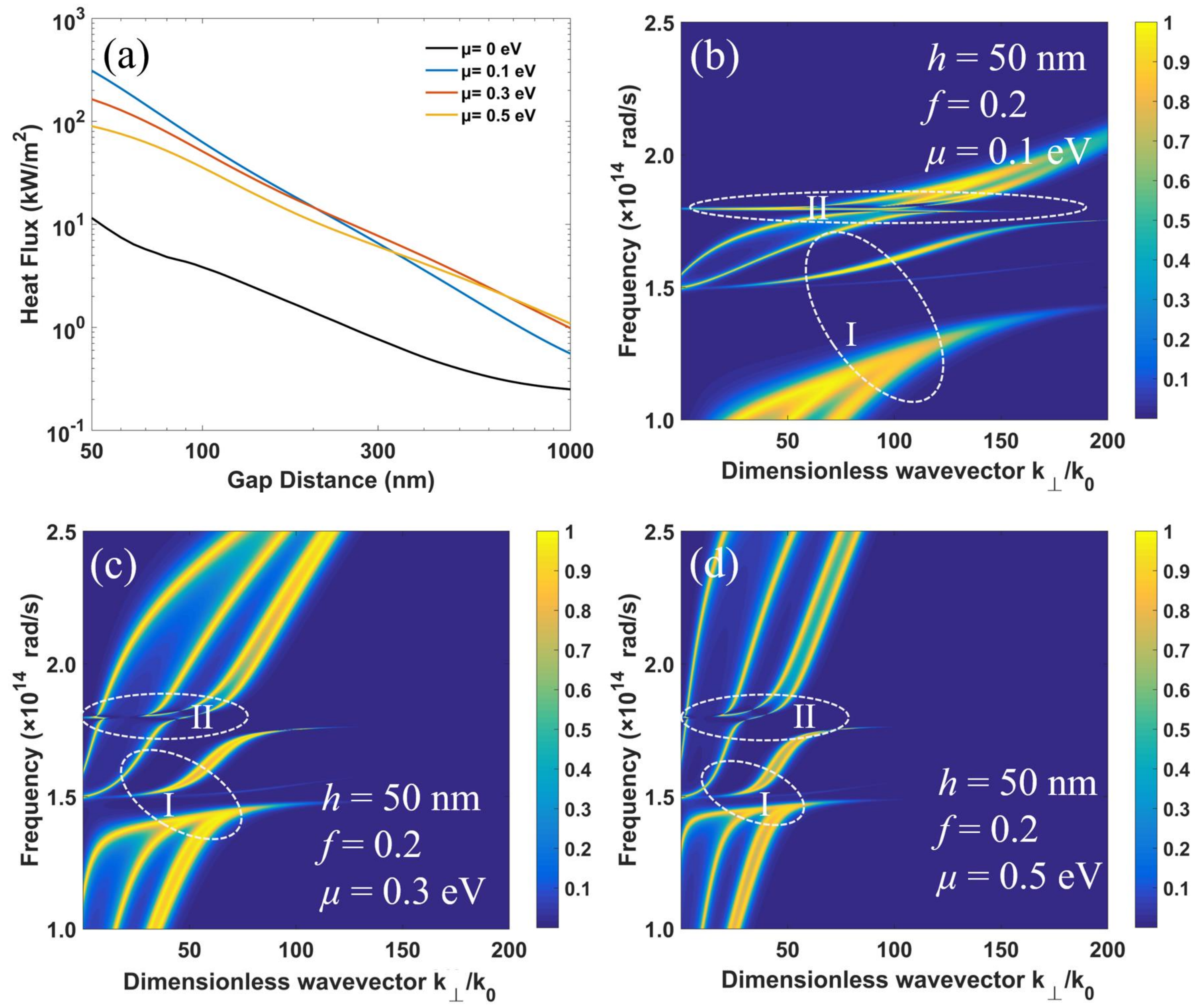}
	\caption{Heat flux versus vacuum gap distance $d$ from 50 to 1000 nm in graphene/SiC nanowire array composite structures with various chemical potentials.}
   \end{figure}

   \begin{figure}[htbp]
	\centering
	\includegraphics[width=8.6cm]{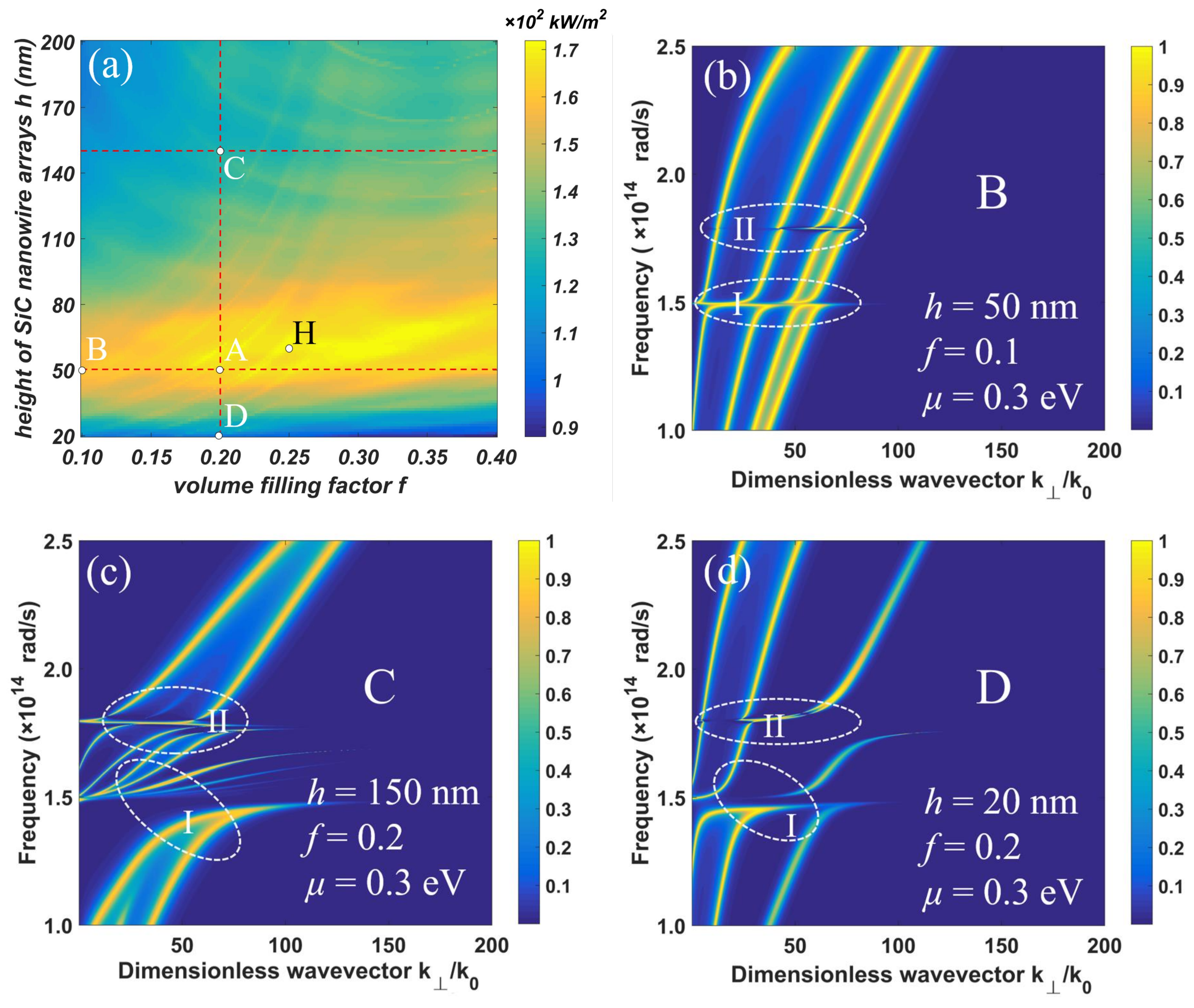}
	\caption{(a) Heat flux between graphene/SiC nanowire arrays composite structures with different volume filling factors and heights of SiC nanowire arrays. Contours of energy transmission coefficients for graphene/SiC nanowire arrays with (b) $f$ = 0.1, $h$ = 50 nm, (c) $f$ = 0.2, $h$ = 150 nm, and (d) $f$ = 0.2, $h$ = 20 nm.}
   \end{figure}

	To further analyze the influence of the volume filling factor and the height of the SiC nanowire arrays on the NFHT, the energy transmission coefficients for points A, B, C, and D are presented in Fig. 5c and Fig. 6b-d. The heat flux of the four points is A $>$ B $>$ C $>$ D. When the volume filling factor of the SiC nanowire arrays changes from 0.2 to 0.1, the HPPs excited region diverges from the SPPs excited region; thus, the SPP-HPP modes’ coupling strength is weak, and the SPPs become the dominant factor in NFHT enhancement, as shown in Fig. 6b. Comparing Fig. 5c and Fig. 6c, more higher-order FRMs of HPPs appear when the height of the SiC nanowire arrays changes from 50 to 150 nm. However, the strong coupling in region I of Fig. 6c becomes weaker. One reason might be that if the SiC nanowire arrays are higher, more FRMs of HPPs will couple or interact with each other and thus decrease the strong coupling effects between SPP and HPP modes. Similarly, the coupling effects in region II of Fig. 5c with shorter SiC nanowire arrays are stronger than those in region II of Fig. 6b, which is understandable because SPPs can be interpreted as a strong localized field on the interface of composite structure that the field intensity evanescently decays away from the interface. However, when the height of the SiC nanowire arrays is 20 nm, only three FRMs are excited in in-plane Reststrahlen band and the strong coupling effects in region I and region II of Fig. 6d are weaker compared with Fig. 5c. Thus, there exists an optimal height of SiC nanowire arrays for strong coupling effects between HPPs and SPPs due to the symmetry mismatch in surface polaritonic modes' coupling.
	
	\section{\label{sec:level4}  Conclusions}
	In this work, we demonstrate that thermally excited SPPs, SPhPs, and HPPs in composite structures lead to a significant enhancement in the NFHT and investigate the role of the strong coupling effects among polaritonic modes. By calculating the energy transmission coefficients and obtaining the near-field dispersion relations, we find that the dispersion relations of composite nanostructures differ substantially from those of isolated graphene, SiC films, and SiC nanowire arrays due to the strong coupling effects between various polaritonic modes. We further identify four pairs of strongly coupled polaritonic modes with considerable Rabi frequencies in the graphene/SiC composite structures, which correspond to four respective FRMs of SPhPs. We confirm that the strong coupling effects between SPPs and HPPs differ between the two Reststrahlen bands due to the alternative types of hyperbolicity. In addition, the effective tunability of the NFHT of graphene/SiC nanowire arrays composite structures is analyzed by adjusting the chemical potential of graphene. The lower the chemical potential of graphene, the higher the NFHT. Moreover, we find that an optimal height exists for graphene/SiC nanowire arrays composite structures for strong coupling effects between HPPs and SPPs at a certain volume filling factor, which can contribute to further enhancement of the NFHT. Due to the limited change in the volume filling factor in this work, the volume filling factor of SiC nanowire arrays has little influence on the NFHT. Therefore, the performance of near-field systems can be effectively controlled by optimizing the parameters of composite structures, which can directly influence the strong coupling effects. This work provides a method to manipulate the near-field heat transfer with the use of strongly coupled surface polaritonic modes.
	
	\section*{Acknowledgments}
	This work is supported by the National Natural Science Foundation of China (Grant No. 51636004), Shanghai Key Fundamental Research Grant (Grant Nos. 18JC1413300, 16JC1403200), National Postdoctoral Program for Innovative Talents of China Postdoctoral Science Foundation (Grant No. BX20180187), and the Foundation for Innovative Research Groups of the National Natural Science Foundation of China (Grant No. 51521004).
	
	\section*{Appendix}
	\emph{Rigorous coupled-wave analysis (RCWA).} The basic structure of graphene/SiC nanowire arrays composite structure is shown in Fig. 7. The layer of SiC nanowire arrays is a mixture of air and SiC nanowires. The radius and period of the SiC nanowires is $R$ and $a$, respectively. Therefore, the volume filling factor can be calculated by $f=\pi R^2/a^2$.
	
    \begin{figure}[htbp]
		\centering
		\includegraphics[width=4cm]{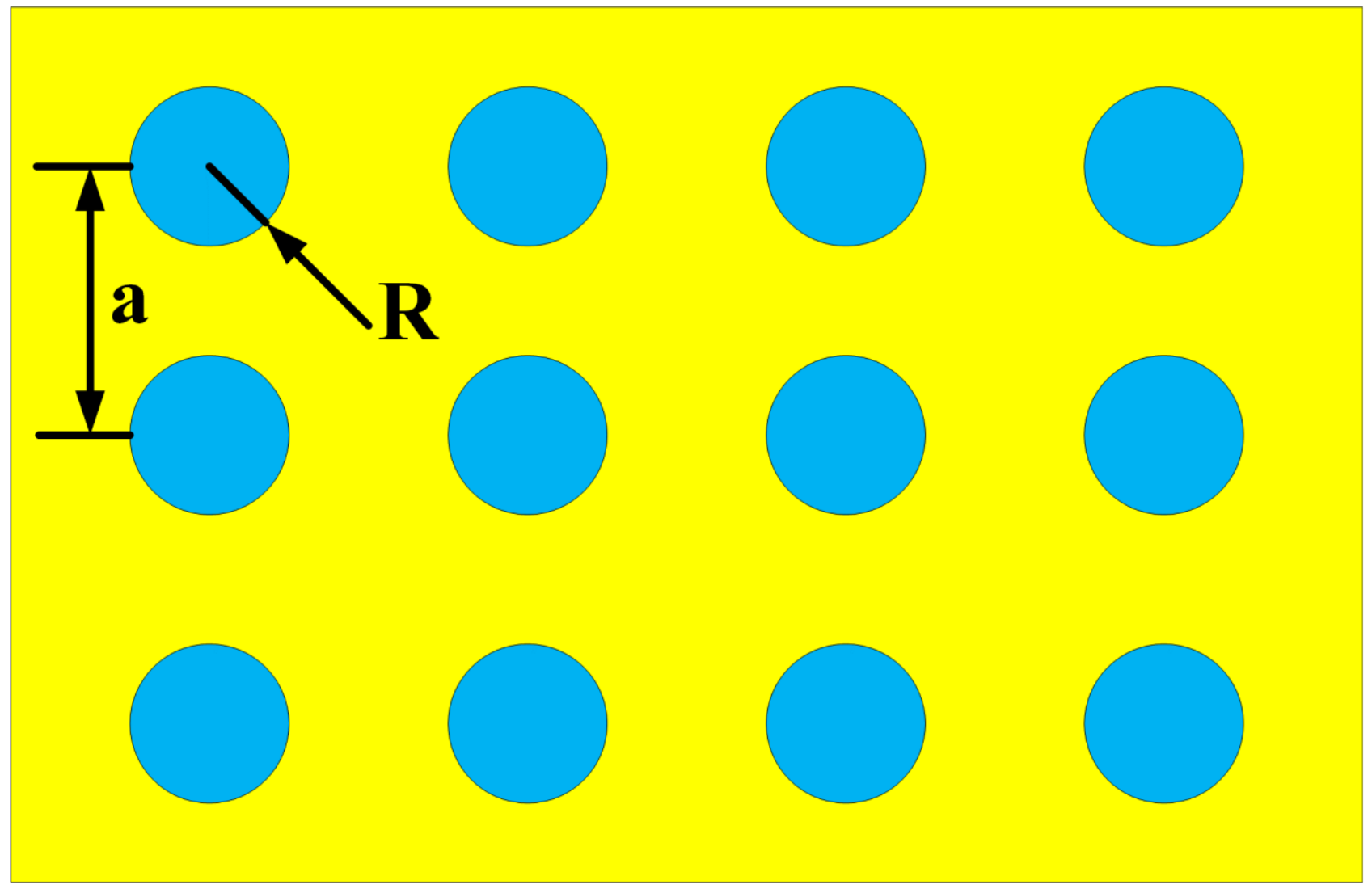}
		\caption{Schematic of the graphene/SiC nanowire arrays composite structure of top view.}
    \end{figure}
	
   The NFHT between graphene/SiC nanowire arrays composite structures based on the exact scattering theory is expressed as \cite{bimonte2009scattering}:
   \begin{equation}
   \begin{split}
   \begin{array}{l}Q=\frac1{8\pi^2}\int_0^\infty\left[\Theta\left(T_E,\omega\right)-\Theta\left(T_R,\omega\right)\right]\operatorname d\omega\\\;\;\;\;\;\;\times\int_{-\mathrm\pi/\mathrm a}^{\mathrm\pi/\mathrm a}\int_{-\infty}^\infty\xi\left(\omega,k_x,k_y\right)\operatorname dk_x\operatorname dk_y\end{array},
   \end{split}
   \end{equation}
   where $\xi\left(\omega,k_x,k_y\right)$ is the sum over polarizations of the transmission probability of the electromagnetic wave. $\left(k_x,k_y\right)$ is the wavevector parallel to the surface planes. Within the RCWA approach, we express the fields in our periodic system as a sum of plane waves using the Bloch theorem. And graphene is modeled as a layer with thickness $\Delta=0.3$ nm with an effective dielectric function $\varepsilon_{graphene}=1+i\sigma/\left(\omega\Delta\right)$ \cite{vakil2011transformation}. Thus, the energy transmission coefficient above can be obtained by combining scattering matrices of the different interfaces in reciprocal space. The energy transmission coefficient can be expressed as \cite{bimonte2009scattering}:   	
   \begin{equation}
    \xi\left(\omega,k_x,k_y\right)=Tr\left\{DW_1D^\dagger W_2\right\},
   \end{equation}	
	where
   \begin{equation}
    D=\left(1-S_1S_2\right)^{-1},   
   \end{equation}	
   \begin{equation}
    W_1={\textstyle\sum_{-1}^{pw}}-S_1{\textstyle\sum_{-1}^{pw}}S_1^\dagger+S_1{\textstyle{\scriptstyle\sum}_{-1}^{ew}}-{\textstyle{\scriptstyle\sum}_{-1}^{ew}}S_1^\dagger,
   \end{equation}
   \begin{equation}
    W_2={\textstyle\sum_{+1}^{pw}}-S_2^\dagger{\textstyle\sum_{+1}^{pw}}S_2+S_2^\dagger{\textstyle{\scriptstyle\sum}_{+1}^{ew}}-{\textstyle{\scriptstyle\sum}_{+1}^{ew}}S_2.
   \end{equation}
   
   Here, $S_1=R_1$ and $S_2=e^{ik_zd}R_2e^{ik_zd}$, where $R_1$ and $R_2$ are the reflection matrices of the two composite structures. These matrices were computed with the scattering-matrix approach \cite{caballero2012generalized}. Moreover, the matrix $\textstyle\sum_{-1\left(+1\right)}^{pw\left(ew\right)}$ is a projector into the propagating and evanescent sector. All these matrices are $\textstyle2N_g\times2N_g$ matrices, where $\textstyle N_g$ is the number of reciprocal lattice vectors included in the plane-wave expansions. In Fig. 8, we show the NFHT as a function for two graphene/SiC nanowire arrays composite structures with $a=100$ nm and $f=0.2$. Notice that both methods yield essentially identical results with less than 0.5\% in the predicted NFHT of the graphene/SiC nanowire arrays composite structures when the gap distance is over 50 nm. However, EMT may fail to predict the NFHT when the gap distance is less than 50 nm. 
   \begin{figure}[htbp]
   	\centering
   	\includegraphics[width=6cm]{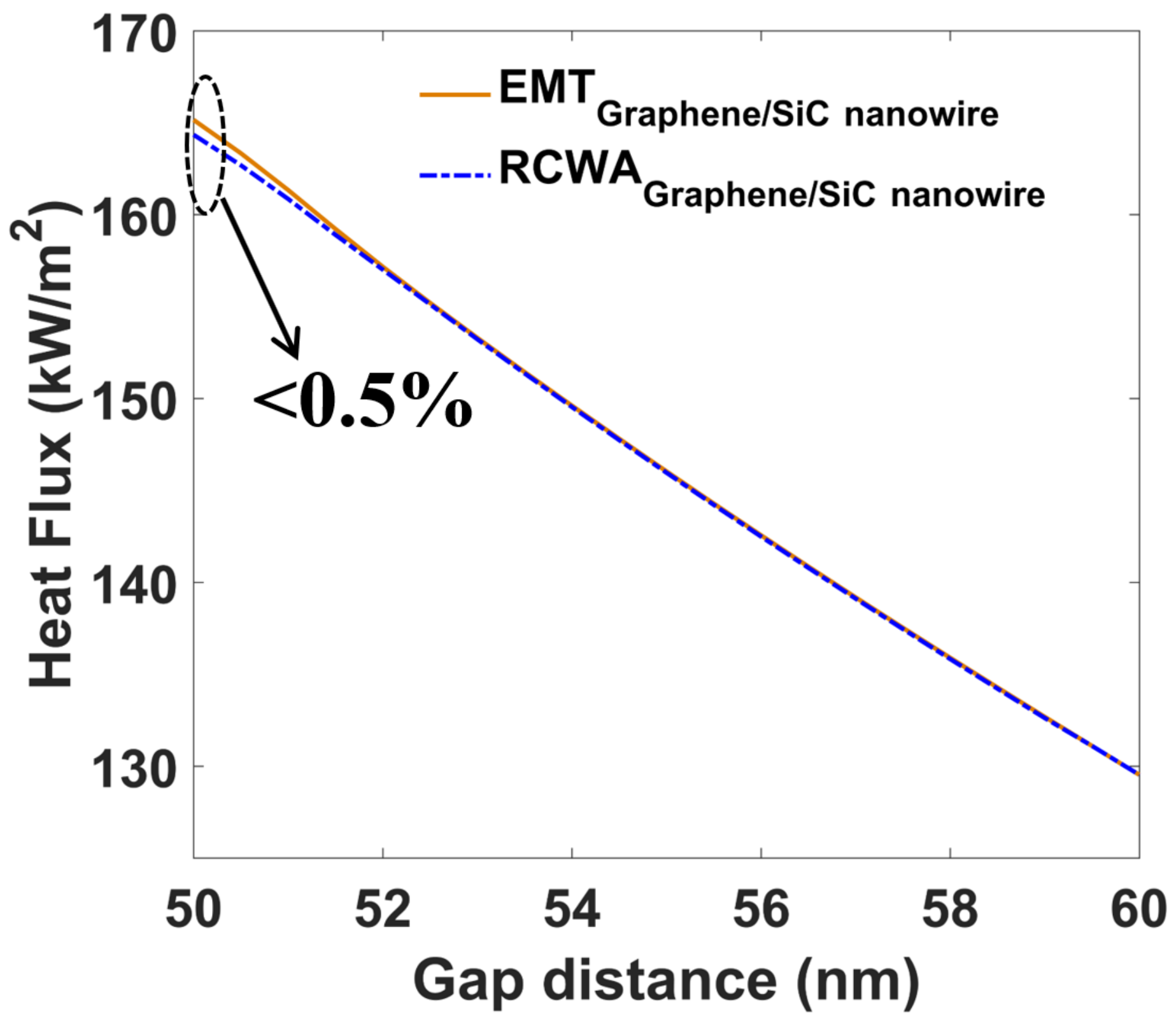}
   	\caption{Heat flux versus vacuum gap distance $d$ from 50 to 60 nm with methods of EMT and RCWA.}
   \end{figure}

   \bibliography{P}
\end{document}